\documentclass[showpacs,prl,twocolumn,amsmath,amssymb,superscriptaddress,amssymb]{revtex4}
\usepackage{graphicx}
\usepackage{dcolumn}
\usepackage{bm}
\usepackage{color}
\usepackage{amssymb}

\begin{document}

\title{Critical correlations in an ultracold Bose gas revealed by means of a temporal Talbot-Lau interferometer}
\author{Wei Xiong}
\affiliation{School of Electronics Engineering and Computer Science, Peking University, Beijing 100871, China}
\author{Xiaoji Zhou}
\email{xjzhou@pku.edu.cn}
\affiliation{School of Electronics Engineering and Computer Science, Peking University, Beijing 100871, China}
\author{Xuguang Yue}
\affiliation{School of Electronics Engineering and Computer Science, Peking University, Beijing 100871, China}
\author{Xuzong Chen}
\affiliation{School of Electronics Engineering and Computer Science, Peking University, Beijing 100871, China}
\author{Biao Wu}
\affiliation{International Center for Quantum Materials, Peking University, 100871, Beijing, China}
\author{Hongwei Xiong}
\email{xionghongwei@wipm.ac.cn}
\affiliation{State Key Laboratory of Magnetic Resonance and Atomic and Molecular Physics, Wuhan Institute of Physics and Mathematics, Chinese Academy of Sciences, Wuhan 430071, China}
\affiliation{Department of Applied Physics, Zhejiang University of Technology, Hangzhou 310023, China}
\date{\today}

\begin{abstract}
We study experimentally the critical correlation in an ultra-cold Bose gas with a
temporal Talbot-Lau (TL) interferometer. Near the critical temperature,  we observe
a bi-modal density distribution in an ultra-cold Bose gas after the application of the TL interferometer.  The measured fraction of the narrower peak in the density distribution displays
a clear peak within the critical regime. The peak position agrees with
the critical temperature calculated with the finite-size and interaction corrections.
The critical exponents are extracted from the peak and they agree
with the critical exponents for the correlation length.
\end{abstract}

.

\maketitle
Near the second-order phase transition \cite{justin,lubensky,privman} from a normal fluid to a superfluid
characterized by a complex order parameter, the diverging spatial correlation length around the critical
temperature is the driving force behind various critical phenomena. There have been enormous
experimental efforts to study the critical phase transition to superfluid in liquid helium \cite{ahlers,ahlers2,lipa}.
One of the endeavors was even carried out in space to get rid of the deviation caused by gravity \cite{lipa2}.
Many critical phenomena, which had long eluded direct experimental study,  have now been studied
experimentally with ultra-cold Bose \cite{greiner,simon,cheng1,cheng2} and Fermi gases \cite{martin}.
In particular, the second-order phase transition from a normal state to a superfluid state has been studied  in
a landmark experiment \cite{donner}, where the critical behavior in the correlation length was revealed
by detecting the interference of two released atomic clouds. This experiment demonstrates that
the spatial correlation in a system can enhance the interference effect, which in turn can be
used to  detect spatial correlation.  Recently, there are also intensive theoretical studies on
the critical behavior with cold atoms~\cite{Kato2008, Campostrini2009, Zhou2010, Diehl2010, Hazzard2011, Fang2011}.

In this Letter we study experimentally the spatial correlation in an ultra-cold Bose gas
with a temporal Talbot-Lau (TL) interferometer.  In such an interferometer, the interference effect
between the correlated atoms reported in Ref.~\cite{donner}  is greatly enhanced because all  atoms in the
Bose gas are involved. With this  TL interferometer, we explore how the spatial correlation in an ultra-cold atomic gas
varies as it undergoes the phase transition from a thermal gas to a Bose-Einstein condensate (BEC). At the critical temperature, we observe a bi-modal density distribution after the application of the TL interferometer.
When the fraction of the narrower density distribution is measured, we find a clear peak
in the critical regime and the peak position agrees well with
the critical temperature calculated for this Bose gas by including the finite-size
and interaction effects.  We have also extracted the critical exponents from the
measured peak and they are very close to the theoretical critical exponents for correlation
length, suggesting that the fraction measured in our experiment is proportional to the
correlation length.

\begin{figure}[ht]
\centerline{\includegraphics[width=.45\textwidth]{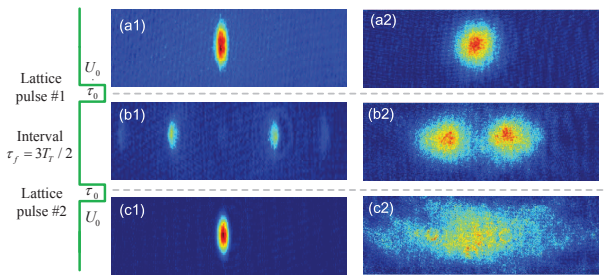}}
\caption{(color online)
Effects on a pure condensate and a thermal gas by the Talbot-Lau interferometer, where
two optical lattice pulses are separated by an interval $\tau_f$ equal to the odd times of
half a Talbot time. The two-dimensional experimental TOF images are  for (a1) a pure condensate
and (a2) a thermal gas before the first lattice pulse; the momentum distribution of (b1) a BEC
and (b2) a thermal gas after the first pulse; the momentum distribution of (c1) a BEC and (c2)
a thermal gas after the second pulse.}
\label{fig1}
\end{figure}

\textit{The temporal TL interferometer for ultra-cold atoms.}---The TL interferometer
has been widely used to reveal the interference property for different systems \cite{cronin}.
Here we use a temporal TL interferometer \cite{cahn,deng} to experimentally study the phase
transition to a BEC. Our TL interferometer (Fig.~\ref{fig1}) consists of two optical lattice
pulses separated by a time interval $\tau_{f}$ equal to the odd times of half a Talbot time
$T_{T}=m\lambda^{2}/2h$  \cite{cahn,deng} with $\lambda$ being the laser wavelength, $m$ the atomic mass and $h$ the Planck's constant. In our experiment, the wavelength $\lambda = 852$ nm, the interval $\tau_f = 3T_T /2$, the pulse width $\tau_0 = 3 \mu$s,
and the lattice depth $U_0 = 80E_R$ with $E_R$ being the recoil energy
of an atom absorbing one lattice photon. The lattice depth is calibrated experimentally
by the Kapitza-Dirac scattering.
The ultra-cold atoms of $^{87}$Rb were prepared in
a magnetic trap with axial frequency 20 Hz and radial frequency
220 Hz \cite{xiong}. The TL interferometer was applied onto the ultra-cold atomic gas
along the axial direction. The time-of-flight (TOF) images are obtained with
the standard absorption imaging method after switching off the magnetic
field and 30 ms free expansion.

During the first lattice pulse, the atomic gas is divided into hundreds of disk-shaped atomic
gases that are physically separated due to the strength of the pulse.  After the pulse, these
disk-shaped atomic gases start to merge and interfere for a given time. The second lattice
pulse again divides it into hundreds of disk-shaped atomic gases, leading to a second interference.
For an initial atomic gas without correlation, the interference is in fact for each individual atom
itself. The overall interference pattern is a simple addition of the interference intensities for all
the individual atoms in the gas. Such a simple addition does not apply for an atomic gas with
spatial correlation larger than the spatial period of the pulsed lattice. Due to the correlation,
one should instead sum up first the interference wave functions for the correlated atoms within
the correlation length,  then square the amplitude of the summed  wave function to
obtain an enhanced interference. This enhancement suggests that the
TL interferometer would be powerful for detecting the spatial correlation among atoms.

Our TL interferometer is effectively an interferometer for atoms around zero
momentum and with spatial correlation larger than the spatial period
of the pulsed optical lattices. To clearly see this, we turn to the momentum space,
where the experimental observation is made,  and focus on atoms
with momenta very close to zero. For these atoms,  after the first optical lattice pulse,
significant proportion of atoms will be transferred to the momentum
around $\pm2n\hbar k$ ($k=2\pi/\lambda$, $n=1, 2, 3, \cdots$) with almost no atoms left
at zero momentum;  after the second optical lattice pulse, the atoms
with momentum around $\pm2n\hbar k$ will be brought back to the
momentum around $0\hbar k$ \cite{clark}. This effect is verified and illustrated
experimentally with  a pure condensate in the left column of Fig. \ref{fig1}.

With this in mind, it is straightforward to see how our interferometer will affect a thermal
atomic cloud without correlation: a very small fraction of atoms that are in the
state of momentum zero are similarly transferred away and brought back to
momentum zero by our interferometer; for majority of the atoms with non-zero momenta,
they will not be brought back to their original momentum states. The end result is an
atomic cloud with much wider distribution in momentum space (see the right
column of Fig. \ref{fig1}). For thermal atomic cloud, because of the lack of correlation between different momentum states, it is expected that there is a trivial mapping before and after the TL interferometer. Both experiment and theoretical simulation show that the momentum distribution is still a Gaussian distribution after the TL interferometer.

\begin{figure*}[ht]
\centerline{\includegraphics[width=1.0\textwidth]{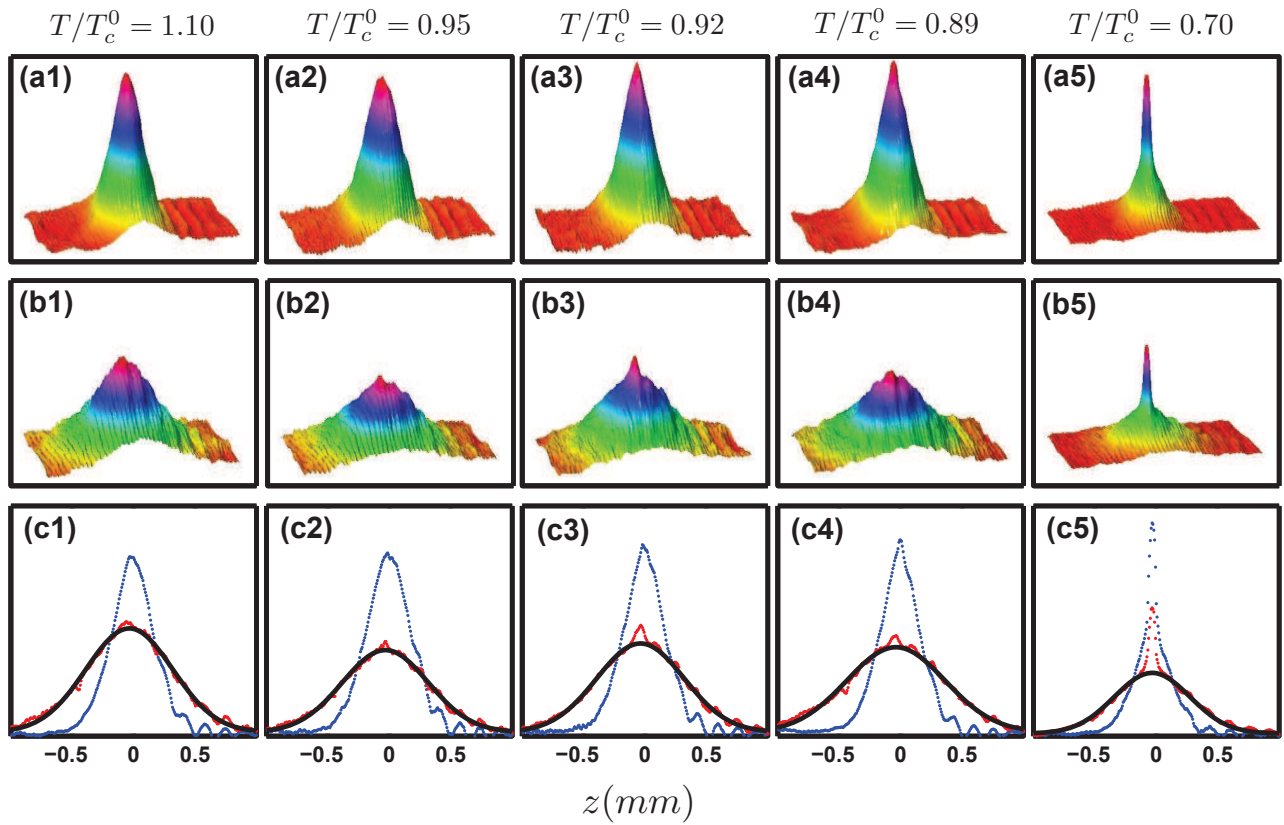}}
\caption{(color online) Revelation of the critical phase transition from a thermal cloud to a BEC by the TL interferometer. The first row shows the density distributions
of the atomic gases before the TL interferometer for five different
temperatures $T/T_{c}^{0}=1.10,0.95,0.92,0.89,0.70$.
The density distributions after the TL interferometer are
in the second row. The dotted blue lines and dotted red lines in (c1)-(c5)
give the one-dimensional density distributions after the integration
of the two-dimensional density distributions along the vertical direction
for (a1)-(a5) and (b1)-(b5), respectively.
A bi-modal structure emerges after the TL pulses
in (b3) and reveals the critical phase transition occurring
around $T/T_{c}^{0}=0.92$. In (c1)-(c5), the solid black lines give the Gaussian fitting to the broad peak of the density distribution after the TL interferometer.}
\label{fig2}
\end{figure*}

Far below the critical temperature which comprises both condensate and thermal atoms, after the application of the TL interferometer, we expect a bi-modal density distribution because the condensate
will still lead to a narrower central peak as  experimentally shown in Fig. \ref{fig2}(b5).

When the system is close to the critical temperature,  there is strong spatial
correlation between atoms, which will enhance the interference as reported in Ref.\cite{donner}.
As our TL interferometer is essentially an interferometer for atoms around momentum zero, this enhancement is expected to increase the population around momentum zero,
causing the momentum distribution around zero to deviate from the Gaussian
distribution after the application of the TL interferometer. Such a deviation from the Gaussian should be bigger for stronger correlation. In a sense, the existence of an order parameter for the correlated atoms within the critical regime makes these atoms have similar behavior to a condensate after the application of the TL interferometer. This physical picture will lead to a bi-modal density distribution after the application of the TL interferometer, similarly to the system far below the critical temperature.

Within the critical regime marked by the Ginzburg temperature $T_G$, i.e. $|T-T_c|<T_G$, the
correlation length diverges near the critical temperature $T_c$ as~\cite{justin,lubensky,privman}
\begin{eqnarray}
\label{length1}
\xi&=&\xi_0/(T-T_c)^\nu\quad{\rm for}\quad T>T_c\,,\\
\xi_T&=&\xi_0^T/(T_c-T)^{\nu_T}\quad{\rm for}\quad T<T_c\,,
\label{length2}
\end{eqnarray}
where the critical exponents $\nu=\nu_T=0.67$. This is for the infinite system in the thermodynamic
limit.  For the finite system in our experiment,  the correlation length no longer diverges. Nevertheless,
the correlation is still the strongest at the critical temperature $T_c$ and the correlation length should have
a peak around  $T_c$. Within the critical regime, this implies that the deviation from the Gaussian distribution after the TL
interferometer is the largest at $T_c$.

\textit{Observation of the critical phase transition.}---With the above expectations,
we have surveyed the ultra-cold atomic gases with the TL interferometer
over a wide range of temperatures with particular attention paid to
the range where the critical phase transition from a thermal cloud
to a BEC occurs. The results for five typical temperatures are shown
in Fig. \ref{fig2}. The system temperature is expressed in terms of
$T_{c}^{0}=0.94\hbar\omega_{ho}N^{1/3}/k_{B}$, the critical
temperature of the corresponding ideal Bose gas without the finite-size
correction \cite{pethick}. $\omega_{ho}=(\omega_{x}\omega_{y}\omega_{z})^{1/3}$
is the geometric average of the harmonic trapping frequencies and $N$ is the total particle number.

Shown in the first row of the figure are TOF images before
the application of  the TL interferometer.  The corresponding
integrated density distributions are shown as dotted blue lines in the third
row. There is a clear bi-modal distribution at temperature $T/T_{c}^{0}=0.7$
(Fig. \ref{fig2}(a5)) and a minor bi-modal distribution at temperature
$T/T_{c}^{0}=0.89$ (Fig. \ref{fig2}(a4)). There are no obvious condensate
fraction at three other higher temperatures.  However, after the application of the TL interferometer, a clear bi-modal (i.e., non-Gaussian) distribution emerges at
$T/T_{c}^{0}=0.92$ (Fig. \ref{fig2}(b3)). In contrast, at a temperature well above the critical temperature ($T/T_{c}^{0}=1.1$), and temperatures slightly lower
($T/T_{c}^{0}=0.89$) and slightly above ($T/T_{c}^{0}=0.95$),
there are no obvious bi-modal structures after the application of the TL interferometer.
This shows that the emergence of the bi-modal distribution at $T/T_{c}^{0}=0.92$
is the result of the diverging  correlation length near the critical
temperature.

To quantify the observed critical behavior, we use the fraction of atoms in
the central peak of the bi-modal structure after the TL pulses for different temperatures.
For each density distribution obtained after the TL interferometer, we fit it along
the direction of the TL pulses with a bi-Gaussian function, and compute the
fraction of the central narrow peak with $f_{r}=1-A_{b}/A_{T}$,
where $A_{b}$ is the area under the broad peak and $A_{T}$ is the
total area in the bi-modal structure. Note that it is an assumption that
the non-Gaussian distribution of momentum observed near $T_c$ is bi-modal.
This assumption is reasonable as indicated in Fig. \ref{fig2}(b3)
(see Supplemental Material for further discussions).

\begin{figure}[ht]
\centerline{\includegraphics[width=.5\textwidth]{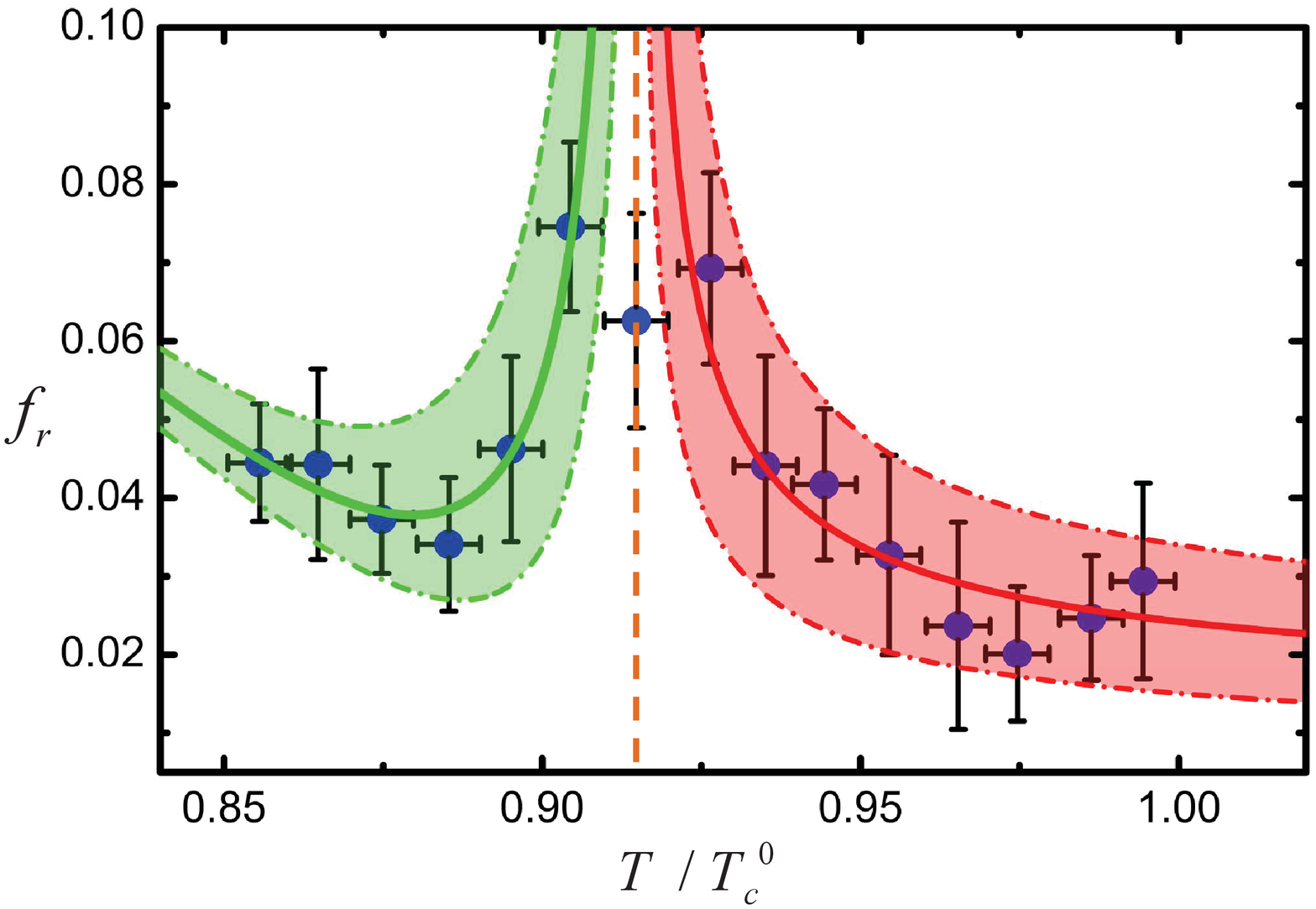}}
\caption{(color online) The measured fraction $f_r$ of the central peak as a function of temperature.
Each point is an average over about ten experimental data within a temperature
step $\delta T/T_{c}^{0}= 0.01$. The two solid lines
are fittings to the data near the peak  with Eqs. (1) and (2). The colored areas represent the uncertainties with a 95\% confidence. The vertical dash line shows the location of the theoretical critical temperature.}
\label{fig3}
\end{figure}

In Fig. \ref{fig3}, we show how this fraction $f_{r}$ changes with
temperature.  Near the critical temperature, a clear peak around $T/T_{c}^{0}=0.92$ is seen.
We also notice a small dip in the peak, which may due to the uncertainty of the
temperature calibration which is about $0.01 T_{c}^{0}$. Because the central
peak after the TL interferometer physically originates
from the spatial interference of the atoms, within the critical regime, it is expected
that the larger the correlation length, the larger the fraction $f_r$.
Hence, this peak implies the critical phase transition from a thermal cloud to a BEC.

As the correlation length diverges at the critical temperature, this peak position
can be regarded as the critical temperature of the system.
Above this critical temperature, the fraction  $f_{r}$ increases
with decreasing temperature,  very similar to what was observed with the
method of the matter-wave interference \cite{donner}. Below this
critical temperature,
we see $f_{r}$ decreases with decreasing system temperature within
the critical regime.

From the peak in Fig. \ref{fig3}, we are able to
determine the critical temperature $T_c$. We find
$T_{c}/T_{c}^{0}=0.92\pm0.01$, indicating a negative shift from
the critical temperature $T_{c}^{0}$ of the ideal Bose gas.
The negative shift is due to the finite-size and interaction corrections of the
critical temperature. The simultaneous
consideration of the finite-size and interaction corrections to the
critical temperature \cite{giorgini, hxiong} gives a negative shift of about 0.08
agreeing well with our experimental result $T_{c}/T_{c}^{0}=0.92\pm0.01$.
The correction to the critical temperature also agrees with
previous experiments \cite{ensher,gerbier,smith1,smith2}. These agreements
further confirm the observation of the critical phase transition in this work.
It is worth to point out that we have computed the Ginzburg temperature
$T_G/T_{c}^{0}$ \cite{giorgini}. The result is that $|T_G-T_c|/T_{c}^{0}\approx 0.06$,
which agrees with the peak width in Fig. \ref{fig3}.

Blow the critical temperature and within the critical regime,  there is a decrease of $f_r$.
This is because that the fraction $f_r$ has two origins,  the condensate fraction that has infinite
coherent length and the correlated thermal fraction that obeys Eq.(\ref{length2}).
Within the critical regime, the contribution from the correlated thermal fraction
dominates, causing the fraction $f_r$ to decrease. In fact, we measured the  condensate fraction
without  the TL interferometer. The results agree with the theoretical result by taking into account
the finite-size and interaction corrections \cite{giorgini,hxiong,hxiong2}. At $T=0.85T_{c}^{0}$,
just outside the critical regime and where the fraction starts to increase, our experiment shows that the condensate fraction
without interferometer is about 0.08 which is of the order
of $f_r$ in Fig.\ref{fig3}.  We emphasize that this understanding of the whole system consisting of
two parts, condensate and correlated thermal atoms, is phenomenological in the spirit of
two-fluid model for liquid helium.  A more rigorous theory is needed to fully understand the
strongly correlated quantum many-body state just below the critical temperature and
how it responds to an interferometer.

\textit{Discussion and outlook.---}Within the critical regime, it is clear that the fraction $f_r$ is a monotonic function of the correlation length.
It is natural to fit the data on the right side of the peak with
$a/(T/T_{c}^{0}-T_{c}/T_{c}^{0})^{\tilde{\nu}}+b$ and the left side
with $a^{\prime}/(T_{c}/T_{c}^{0}-T/T_{c}^{0})^{\tilde{\nu}^{\prime}}+b^{\prime}+c^{\prime}(T_{c}/T_{c}^{0}-T/T_{c}^{0})$.
The additional term for the left side is to account for the
condensate fraction which appears below the critical temperature. Rigorously speaking, the correlation length should display a smooth evolution within the critical regime due to the finite-size characteristic of our experiment. Similarly to Ref. \cite{donner}, because of the uncertainty of the temperature calibration, we do not discuss the finite-size effect to the correlation length in this work.

From the data fitting, we get
$\tilde{\nu}^{\prime}=0.70\pm0.08$ and $\tilde{\nu}=0.70\pm0.11$.
The uncertainties correspond to a $95\%$ confidence level. These two
critical exponents are very close to $\nu=\nu_{T}=0.67$, the
theoretical value of the correlation length's critical exponent for
the universality class of three-dimensional XY model \cite{burovski, Yukalov2007}.
If we assume that $f_r\sim \xi^\alpha$ and the theoretical value of the critical exponent is exact,
we have $\alpha\approx 1.04\pm 0.12$. This experimental result indicates
that the fraction $f_{r}$ near the peak in Fig. \ref{fig3} is proportional
to the correlation length.

This encouraging indication is supported by a phenomenological theory
(see Supplemental Material).
However, we still do not have a rigorous theory how a strongly correlated system responds to an interferometer as
the  action  of an interferometer  can no longer be regarded as a small perturbation.  This is
different from known measurements on critical systems, for example, the heat capacity measurement
on liquid helium, which can be regarded as a perturbation and have negligible effects on the systems.

In summary,  we have studied the critical behavior of interacting
ultra-cold Bose gases with a temporal TL interferometer.
A peak in the fraction of the narrower peak of the density distribution was clearly
observed across the  phase transition from a thermal gas to a BEC. This experimental study
of  the ultra-cold atomic gas within the critical regime opens the way to relevant theoretical studies
of critical dynamics, which has many open problems \cite{justin,lubensky,privman}. It also gives a
new method to measure the critical temperature.
We expect that the temporal TL interferometer be used to study
a wide range of the critical phase transitions, such as the quantum
phase transition from superfluid to Mott insulator for cold atoms
in an optical lattice \cite{greiner}, ultra-cold Fermi gas \cite{martin} and
the quantum magnetism with the cold atoms and molecules as a quantum simulator \cite{sachdev,jo,simon, Gorshkov2011}.

\begin{acknowledgments}
This work was supported by the NBRP of China (2011CB921500, 2012CB921300),  the NSF of China (61078026,11175246, 10825417),  and the RFDP of China (20110001110091).
\end{acknowledgments}



\end{document}


\global\def\thesection{S.\Roman{section}}\hspace{1em}
\global\def\theequation{S.\arabic{equation}}
\global\long\def\thefigure{S\arabic{figure}}

\noindent\textbf{\large{Supplemental Material}}
\bigskip

\noindent\textbf{Temporal Talbot-Lau interferometer for a condensate.} Our temporal TL
interferometer  has been described in the main text and its effects are illustrated in Fig.~1
of the main text. We here provide a detailed theory and explain why the momentum distribution
of the condensate evolves as shown in the left column of Fig.~1 of the main text under
the TL interferometer.

Consider a condensate with very narrow momentum distribution
around zero. After the first short optical lattice pulse, which is
described as $V(z)=U_{0}\cos^{2}(kz)$, the atoms are in a state approximately
described by \cite{clark}
\begin{eqnarray}
\phi(\tau_{0}) & \propto & \exp\{i\frac{U_{0}\tau_{0}}{\hbar}\cos^{2}(kz)\}\,|0\rangle\nonumber \\
 & \propto & \sum_{n=-\infty}^{\infty}i^{n}J_{n}(U_{0}\tau_{0}/2\hbar)|2n\hbar k\rangle\,,
\end{eqnarray}
where $|2n\hbar k\rangle$ ($n=0, 1, 2, \cdots$) represents a state of momentum $2n\hbar k$ and $J_{n}$ is the Bessel function of the first kind. Our experimental parameters have been chosen such that $J_{0}(U_{0}\tau_{0}/2\hbar)=0$, which means that the zero momentum is no longer occupied after the first pulse (Fig.~1(b1)).

After a free evolution of time $\tau_{f}$, the second pulse
which is identical to the first one is applied. The quantum state
is again transformed and becomes
\begin{eqnarray}
\phi(2\tau_{0}+\tau_{f}) & \propto & \sum_{n=-\infty}^{\infty}i^{n}\big[\sum_{m=-\infty}^{\infty}J_{n-m}(U_{0}\tau_{0}/2\hbar)
e^{-i\theta_{m}}J_{m}(U_{0}\tau_{0}/2\hbar)\big]|2n\hbar k\rangle\,,\label{second}
\end{eqnarray}
where $\theta_{m}=2\pi m^{2}\tau_f/T_{T}$. Since $\tau_{f}=3T_{T}/2$
in our experiment, we have
\begin{equation}
\phi(2\tau_{0}+3T_{T}/2)\propto\sum_{n=-\infty}^{\infty}i^{n}J_{n}(0)|2n\hbar k\rangle
=|0\rangle.\label{back}
\end{equation}
This means that the second pulse has completely brought the system
back to the state of zero momentum (Fig.~1(c1)).

Note that in getting Eq.~(\ref{second}), there is an implicit assumption that during the second optical lattice pulse, there is full spatial overlapping between different momentum state $|2n\hbar k\rangle$. For our experimental parameters for a pure condensate, this condition holds. Also note that in the above derivation,  we have ignored
the non-homogeneous distribution of the condensate in a real experiment. It is clear from
the experimental result in Fig.~1 that this non-homogeneity is not important. This is also verified by our
 numerical simulation.

The effect of our TL interferometer is observed experimentally
with a pure $^{87}$Rb BEC as shown in the left component of Fig.~1, where the experimental
time-of-flight (TOF) images are given. The TOF images are obtained
with the standard absorption imaging method after switching off the magnetic
field and 30 ms free expansion. A cigar-shaped pure condensate of
$10^{5}$ atoms was created with axial frequency $20$ Hz and radial
frequency 220 Hz. Its momentum distribution is shown in Fig.~1(a1).
When only the first pulse is applied along the axial direction, as
shown in Fig.~1(b1), most of the atoms (about $60\%$) are transferred
to the momentum around $\pm2\hbar k$. About $36\%$ of the atoms
are populated around $\pm4\hbar k$ with the population around momentum
zero almost completely depleted. After a time interval of $3T_{T}/2$,
the second pulse with the exact same parameters is applied, as shown
in Fig.~1(c1).

For thermal atomic cloud far away from the critical temperature, because it is an incoherent superposition of different momentum states, after our TL interferometer, we observe a broadened momentum distribution shown on the right of Fig.~1. Our numerical simulation also gives this behavior. \\


\noindent\textbf{Temporal Talbot-Lau interferometer for the system near the critical
temperature.} In our experiment, the TL interferometer is applied along
the axial $z$ direction of a cigar-shaped atomic cloud, and we can
treat it as an effectively one dimensional system. Near the critical
temperature, we can introduce an order parameter $\psi\left(z\right)=\langle \hat{\Psi}^{\dag}\left(z\right)\rangle$ to describe the
system based on the widely used phenomenological theory for phase transition.

The evolution of the order parameter $\psi\left(z,t\right)$ can be studied with the
following propagator method
\begin{equation}
\psi\left(z,t\right)=\int dz_1 K\left(z,t;z_{1},t_{0}\right)\psi\left(z_{1},t_{0}\right).\label{eq:prop}
\end{equation}
Here $K\left(z,t;z_{1},t_{0}\right)$ is the propagator, describing how the system evolves.
For our experiment, one may regard it as the evolution caused by the application of the TL
interferometer.  The density distribution is then
\begin{equation}
n\left(z,t\right)=\psi^{*}\left(z,t\right)\psi\left(z,t\right)=\int K^{*}\left(z,t;z_{1},t_{0}\right)K\left(z,t;z_{2},t_{0}\right)\psi^{*}\left(z_{1},t_{0}\right)
\psi\left(z_{2},t_{0}\right)dz_{1}dz_{2}.
\label{dynorder}
\end{equation}

Within the critical regime, the order parameter has significant amplitude and phase fluctuations. We can write the order parameter as $\psi\left(z,t_0\right)=\psi_0\left(z,t_0\right)\theta\left(z,t_0\right)$. Here  $\psi_0\left(z,t_0\right)=\sqrt{\langle\hat{\Psi}^{\dagger}(z,t_0)\hat{\Psi}(z,t_0)\rangle}$ is a smooth function with uniform phase while $\theta(z,t_0)$ describes the amplitude and phase fluctuations. With this, Eq. (\ref{dynorder}) becomes
\begin{equation}
n\left(z,t\right)=\int K^{*}\left(z,t;z_{1},t_{0}\right)K\left(z,t;z_{2},t_{0}\right)\psi_{0}^{*}\left(z_{1},t_{0}\right)
\theta^{*}\left(z_1,t_0\right)\psi_{0}\left(z_{2},t_{0}\right)\theta\left(z_2,t_0\right)dz_{1}dz_{2}.
\label{dyn}
\end{equation}
We now take the ensemble averaging, which yields $\langle\theta^{*}\left(z_1,t_0\right)\theta\left(z_2,t_0\right)\rangle=G\left(z_1,z_2\right)$, where $G\left(z_1,z_2\right)$ is the correlation function between two points $z_1$ and $z_2$ \cite{justin,lubensky}.  After the ensemble averaging, we obtain the population of the atoms with the
second-order correlation,
\begin{eqnarray}
&  n_0\left(z,t\right)=\int K^{*}\left(z,t;z_{1},t_{0}\right)K\left(z,t;z_{2},t_{0}\right)\psi_0^{*}\left(z_{1},t_{0}\right)
\langle\theta^{*}\left(z_1,t_0\right)\theta\left(z_2,t_0\right)\rangle\psi_0\left(z_{2},t_{0}\right)dz_{1}dz_{2} & \nonumber  \\
  & =\int K^{*}\left(z,t;z_{1},t_{0}\right)K\left(z,t;z_{2},t_{0}\right)\psi_0^{*}\left(z_{1},t_{0}\right)
  G\left(z_1,z_2\right)\psi_0\left(z_{2},t_{0}\right)dz_{1}dz_{2}.  &
\label{Findyn}
\end{eqnarray}
Although the above expression is rather general and applies to almost all systems,  useful information can still
extracted. When the system is far above the critical temperature, we have $G=0$ and $n_0=0$ as expected. When the system is far below the critical temperature, where it is dominated by the condensate fraction, the correlation function is $G(z_1,z_2)=1$. Hence, $n_0$ relates directly to the condensate.  In general, it is clear from the above equation, the population
of the correlated atoms is a monotonic functional of the second-order correlation function $G\left(z_1,z_2\right)$:
when $G$ is strong, the population $n_0$ is larger. This fact can qualitatively explain the experimental
observations in Ref. ~\cite{donner} and our experiment. However, the exact relation is complicated by
the existence of other functions, such as the propagator $K$,  in Eq.(\ref{Findyn}).
As it is difficult to obtain the exact form of $K$ for this strongly correlated system,  the quantitative
explanation of these two experiments is elusive.

To make progress, we write the whole density
distribution $n_{all}\left(z,t\right)$ as
\begin{equation}
n_{all}\left(z,t\right)=n_0\left(z,t\right)+n_{incoherent}\left(z,t\right).
\end{equation}
For these correlated atoms, the width of momentum distribution of $n_0\left(z,t\right)$ is assumed to have
the same as that of a pure condensate with the same form of the order parameter. The incoherent component,
however, has a wider density distribution because of the lack of spatial interference after the second pulsed optical lattice. Hence we expect a bi-modal density distribution within the critical regime as observed in our experiment.
In the main text, $f_r=\int n_0\left(z,t\right)dz/\int n_{all}\left(z,t\right)dz$ is the measured
fraction of the correlated atoms.


For the TL interferometer, the above expression can be further
simplified with a discrete method. After the first optical lattice
pulse, the atomic cloud is broken up into hundreds of subsystems.
If the atoms in the cloud are strongly correlated, each of these subsystems
will contribute an order parameter with a complex amplitude $\psi_{j}$ ($j$ denotes the
$j$th subsystem) to the total amplitude. After the second pulsed optical lattice,
we have at the end of the TL interferometer action
\begin{equation}
n_0\left(z,t\right)=\sum_{ij}\psi_{i}^{*}\left(z,t\right)G_{ij}\psi_{j}\left(z,t\right),
\label{inter}
\end{equation}
where $G_{ij}=G(|z_{i}-z_{j}|)$ is the correlation function between
two points $z_{i}$ and $z_{j}$.

During the interference process, each subsystem $\psi_{j}(z,t)$ will expand and
become almost completely overlap with each other. At the same time, within the
critical regime,  the correlation length $\xi$ in $G(|z_{i}-z_{j}|)$ is much larger than
the spatial period of the pulsed optical lattice while much smaller than the length of the system (although we can not measure directly the value of the correlation length, the work in Ref.~\cite{donner} with similar parameters
shows that the correlation length in the critical regime satisfies this condition).
With all these consideration, we have
\begin{equation}
n_{0}\left(z,t\right)\propto\sum_{ij}\psi_{i}^{*}\left(z,t\right)\psi_{j}\left(z,t\right)
\sum_{ij}G_{ij}\propto\sum_{ij}\psi_{i}^{*}\left(z,t\right)\psi_{j}\left(z,t\right)\int\int dz_{1}dz_{2}G(|z_{1}-z_{2}|)\,.
\label{inter2}
\end{equation}
 Another reason that the above approximation is good lies in that the atoms concentrate in the
 central region of the magnetic trap, thus the homogeneous approximation can be used.

 In the critical regime above the critical temperature, the
correlation function is
\begin{equation}
G(z)\propto\frac{\exp\left(-|z|/\xi\right)}{|z|}.
\label{above}
\end{equation}
In the critical regime below the critical temperature, the correlation
function is dominated by the transverse part and we have \cite{privman}
\begin{equation}
G(z)\propto\frac{\xi_{T}}{|z|}.
\label{below}
\end{equation}
Nevertheless, after a simple integral transform, we have for both
cases
\begin{equation}
n_{0}\propto\xi\,.
\end{equation}
Therefore, we have the fraction of the correlated atoms $f_{r}\propto\xi$
in the critical regime. This simple relation shows
why the TL interferometer has the ability to reveal the critical behavior.

The above analysis relies on two assumptions. (1) The first
pulsed optical lattice will not change the correlation function Eq.
(\ref{above}) or Eq. (\ref{below}) when the correlation length
is larger than the spatial period of the pulsed optical lattice. This
is a valid assumption for the experimental pulse width $\tau_{0}=3\mu$s,
which is much shorter than the thermal equilibrium time. (2) There
is no thermal equilibrium between two pulsed optical lattices in the
TL interferometer. In our experiment, the interval between two pulsed
optical lattices is $\tau_{f}=3T_{T}/2$, which
is much smaller than the thermal equilibrium time of the order of
several milliseconds.

With the above analysis on the correlated
atoms, we see that the TL interferometry has the effect of separating
the correlated atoms and incoherent atoms. Because the correlated
atoms will be brought back to the state near zero momentum after the
TL interferometry, the density distribution of the correlated atoms
leads to the central peak in the whole density distribution. This
leads to the bi-modal structure in the density distribution after
the TL interferometer, if there is significant fraction of correlated
atoms.

For $T/T_{c}^{0}=0.70$, without the application of the TL interferometer, we see a
clear bi-modal structure which shows the density distribution of the condensate and
thermal atoms, respectively. After the application of the TL interferometer, there is
no essential change in the bi-modal structure, because the condensate will still appear
in the central peak in the bi-modal density distribution, while the thermal atoms give
only the wider background. At the critical temperature  $T/T_{c}^{0}=0.92$, however,
the bi-modal structure appears after the application of the TL interferometer. The only
interpretation is that the correlated atoms will appear in the central peak
although there is no condensate in this situation.
